\documentclass[final,5p,times,twocolumn]{elsarticle} 
\usepackage{graphicx}
\usepackage{amsmath}
\usepackage{mathtools}
\usepackage{amssymb}
\usepackage{lineno}
\nolinenumbers
\usepackage[colorinlistoftodos]{todonotes}
\usepackage[colorlinks=true, allcolors=blue]{hyperref}

\usepackage{multirow}
\usepackage{multicol}
\usepackage{comment} 

\hypersetup{
        pdfauthor          = {Pavel Parygin},
        pdftitle           = {SiPM single cell analysis},
        pdfsubject         = {sipm},
        pdfkeywords        = {CMS, physics, SiPM, radiation},
        urlcolor           = red,
        citecolor          = blue,
        linkcolor          = blue,
        pdfborder          = {1 1 1}
        }

\begin{document}
\begin{frontmatter}

\title{Radiation hardness study using SiPMs with single-cell readout}
\author[A]{E.~Garutti}
\author[B]{E.~Popova} 
\author[B]{P.~Parygin}
\author[B]{O.~Bychkova}
\author[C]{A.~Kaminsky}
\author[A]{S.~Martens}
\author[A]{J.~Schwandt}
\author[B]{A.~Stifutkin} 
 
\address[A]{University of Hamburg, Hamburg, Germany}
\address[B]{National Research Nuclear University MEPhI, Moscow, Russia}
\address[C]{Skobeltsyn Institute of Nuclear Physics, Lomonosov Moscow State University, Moscow, Russia}

\date{\today}
\graphicspath{ {fig/} }

\begin{abstract}
A dedicated single-cell SiPM structure is designed and measured to investigate the radiation damage effects on the gain and turn-off voltage of SiPMs exposed to a reactor neutron fluence up to $\Phi$~=~5e13~cm$^{-2}$. The cell has a pitch of 15~$\mu$m. The fluence dependence of gain and turn-off voltage are reported.
A reduction of the gain by 19\% and an increase of $V_{off}$ by $\approx$0.5~V is observed after $\Phi$ = 5e13 cm$^{-2}$.   
\end{abstract}


\end{frontmatter}

\section{Introduction}
Silicon photomultipliers (SiPMs), thanks to their excellent performance, are becoming the photodetectors of choice for many applications. One major limitation, in particular for their use at high-luminosity colliders, is the radiation damage induced by charged or neutral hadrons. 
As SiPMs detect single charge carriers, radiation damage is a major concern when operating these devices in harsh radiation environments (i.e. CMS and LHCb detectors at LHC, detectors at the proposed International Linear Collider (ILC), detectors for space experiments, etc.). 
Results on the operation of irradiated SiPMs with X-ray, gamma, electron, proton and neutron sources are reviewed in~\cite{Garutti_2019}. The most critical effect of radiation on SiPMs is the increase of dark count rate, which makes it impossible to resolve signals generated by a single photon from the noise. Once the single photo-electron (SPE) resolution is lost the SiPM gain cannot be directly determined as the separation of the peaks in a SPE distribution. Additionally, in the absence of single-photon counting capability the turn-off voltage cannot be extracted from the linear relation between the excess bias voltage and the gain, $G=C_{pix}(V_{bias}-V_{off})/q_0$.
In the literature, I-V measurements are used to determine $V^{IV}_{bd}$ in this case. It should be noted that these are not the same quantities, and that a difference between $V^{IV}_{bd}$, the turn-on voltage for the Geiger avalanche (or breakdown voltage), and the turn-off voltage, $V_{off}$ is demonstrated in Ref.~\cite{Chmill:2016msk}. In this paper the difference $V_{ov}=V_{bias}-V_{off}$ is referred to as overvoltage. Measurements of $V^{IV}_{bd}$ are challenging for a single cell and are not presented here, but will be performed in the future.

There are three key questions relevant to understanding of the behavior of irradiated SiPM: 
\begin{enumerate}
    \item Is the value of $V_{off}$ changing with fluence, and if so, is the change correlated to that of $V^{IV}_{bd}$?
    \item Is the SiPM gain, $G$ at fix $V_{ov}$ changing with fluence? 
    \item Is the SiPM photo-detection efficiency, $PDE$ at fix $V_{ov}$ changing with fluence? 
\end{enumerate}

To address these questions, we conceived a dedicated structure with a  metal layer configuration that enabled the readout of one single-cell separately from the others in a SiPM, and Hamamatsu Photonics K.K. (HPK) \cite{HPK} kindly provided a sample with a structure that responded to these needs. 
The structure was produced together with the standard MPPCs of the S14160 series. The possibility to readout a single cell enables the study of highly irradiated SiPMs, retaining the single-photon counting capability. 

In this paper, the dedicated single-cell structure and the setup used for its characterization are introduced in Sec.~\ref{sec:setup}. The data analysis method is discussed in Sec.~\ref{sec:method}. Sec.~\ref{sec:results} presents the results for $G$ and $V_{off}$ measured on single-cell structures irradiated with reactor neutrons up to $\Phi$~=~5e13~cm$^{-2}$. With these measurements question number 2 can be answered. To answer questions 1 and 3 further studies will need to be performed, which require further developments of the measurement setup.

\section{Device and setup description}
\label{sec:setup}
\begin{figure}[ht]
    \centering
    \includegraphics[width=0.4\textwidth]{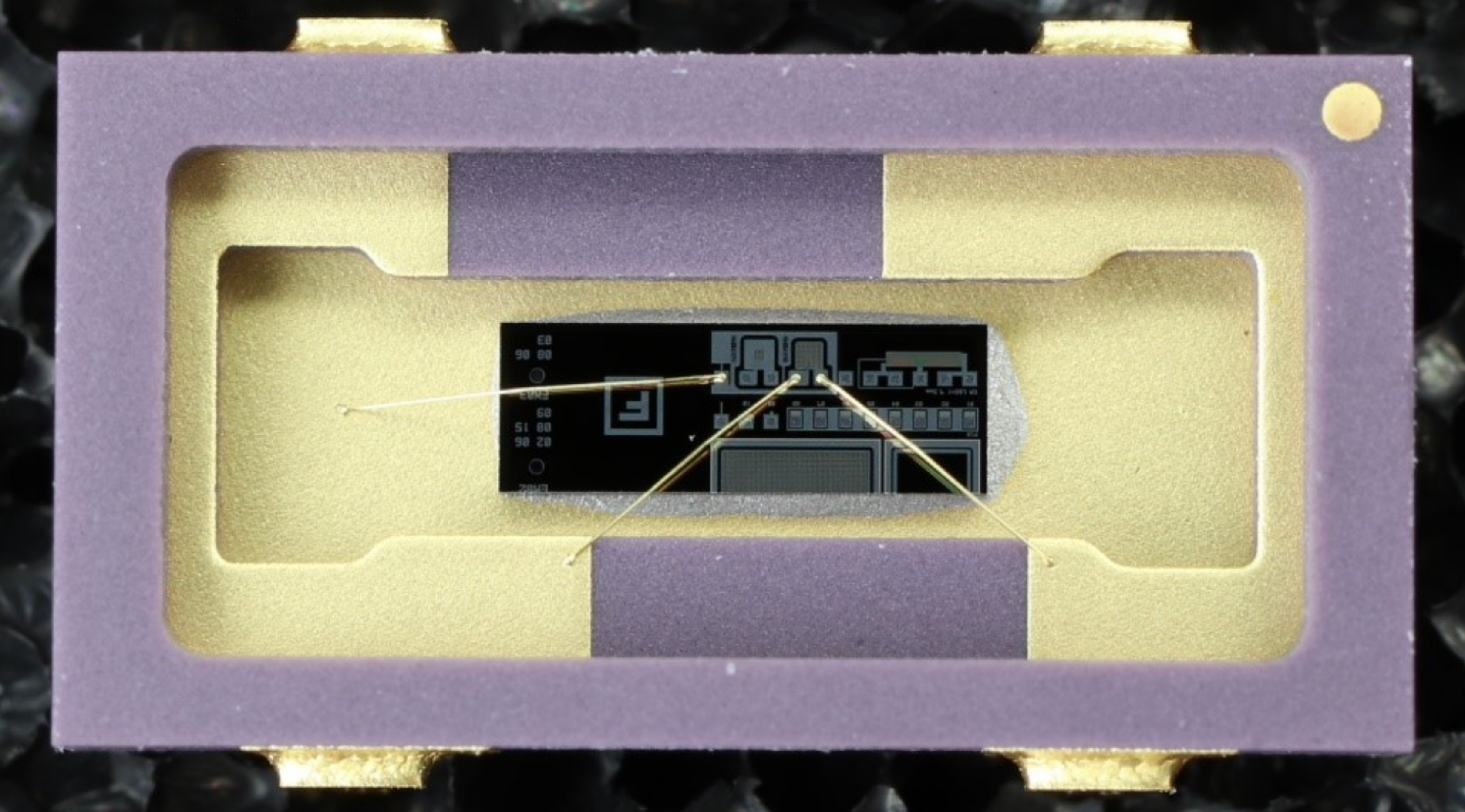}
    \caption{Hamamatsu SiPM test structure of S14160 series.}
    \label{fig:DUTpic1}
\end{figure}

\begin{figure}[ht]
    \centering
    \includegraphics[width=0.45\textwidth]{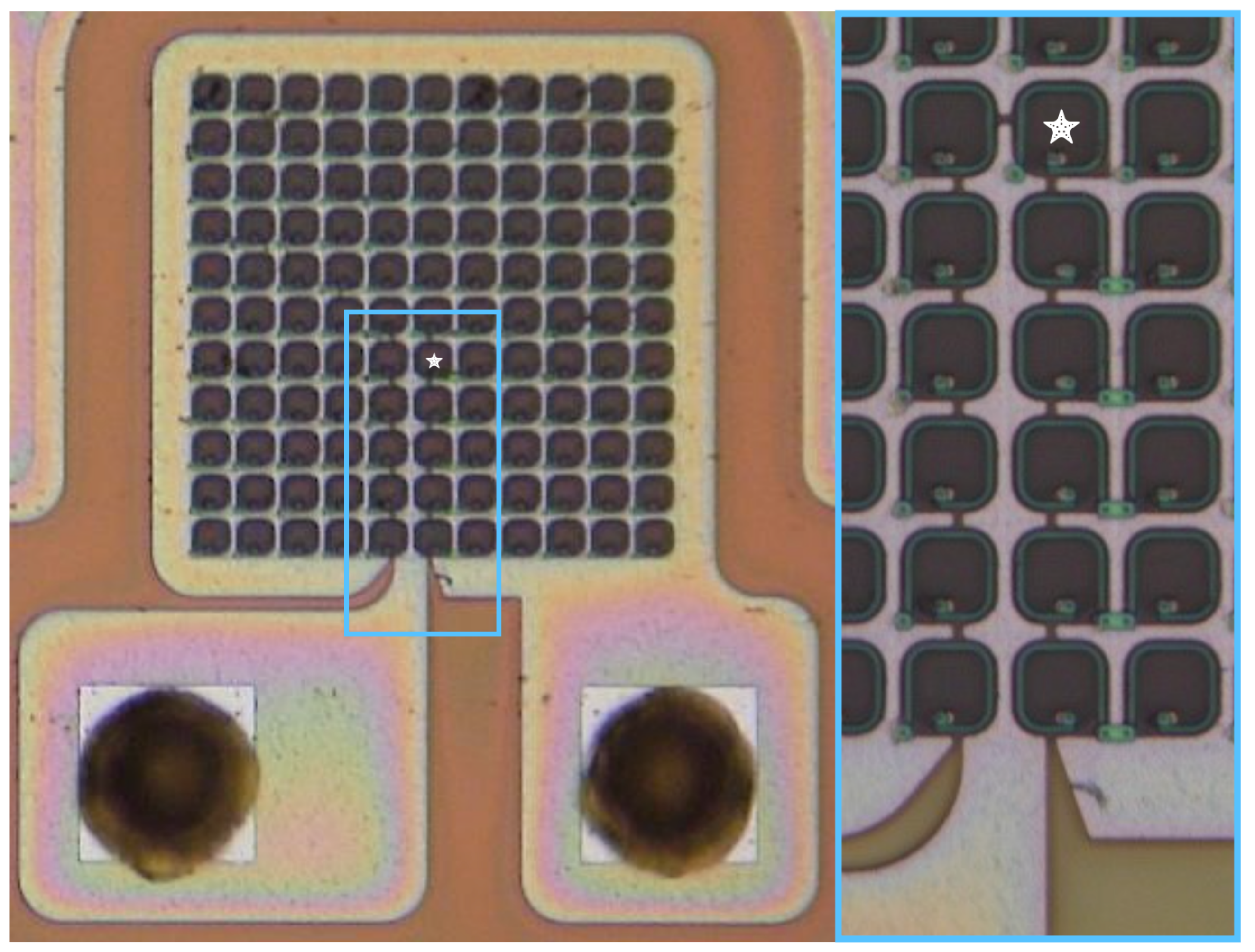}
    \caption{Microscopic image of the studied 11x11-cell array. The disconnected cell is marked with a star symbol. On the 50x-zoomed image on the right-hand side one can see the cell isolation by gaps in the metallization layer.}
    \label{fig:DUTpic2}
\end{figure}

The device under test (DUT) is a Hamamatsu SiPM test structure of S14160 series \cite{HPKS14160} glued on the ceramic package. Given SiPM series has a trenches of 0.5 $\mu$m width between the cells for low crosstalk \cite{hpkCHEF2019}. A picture of the DUT is shown in Figure~\ref{fig:DUTpic1}. It has an array of 11x11 cells with 15 $\mu$m pitch (see Figure~\ref{fig:DUTpic2}). The central cell of the array is disconnected from the others and has its own output contact pad. Therefore 1 cell and 120 cells have a common cathode but separate anodes. 

The setup consists of a climate chamber, a dual-channel bias and readout board with the amplifiers, and an oscilloscope with a sample rate of 10~GS/s for data acquisition.

For the illumination of the device a pulsed 451~nm laser is used, with a 50~ps pulse length. A filter mount placed outside the chamber is used for changing the laser light intensity. The light delivery system consists of two optical fibers, the first with 125~$\mu$m-core and 10~m length from the laser to the filter mount, and the second with 365~$\mu$m-core and 2~m length from the filter mount to the DUT. For the measurements without illumination the optical fiber was blocked in the filter mount by a beam blocker, while the laser was kept on. 

\begin{figure}[ht]
    \centering
    \includegraphics[width=0.9\linewidth]{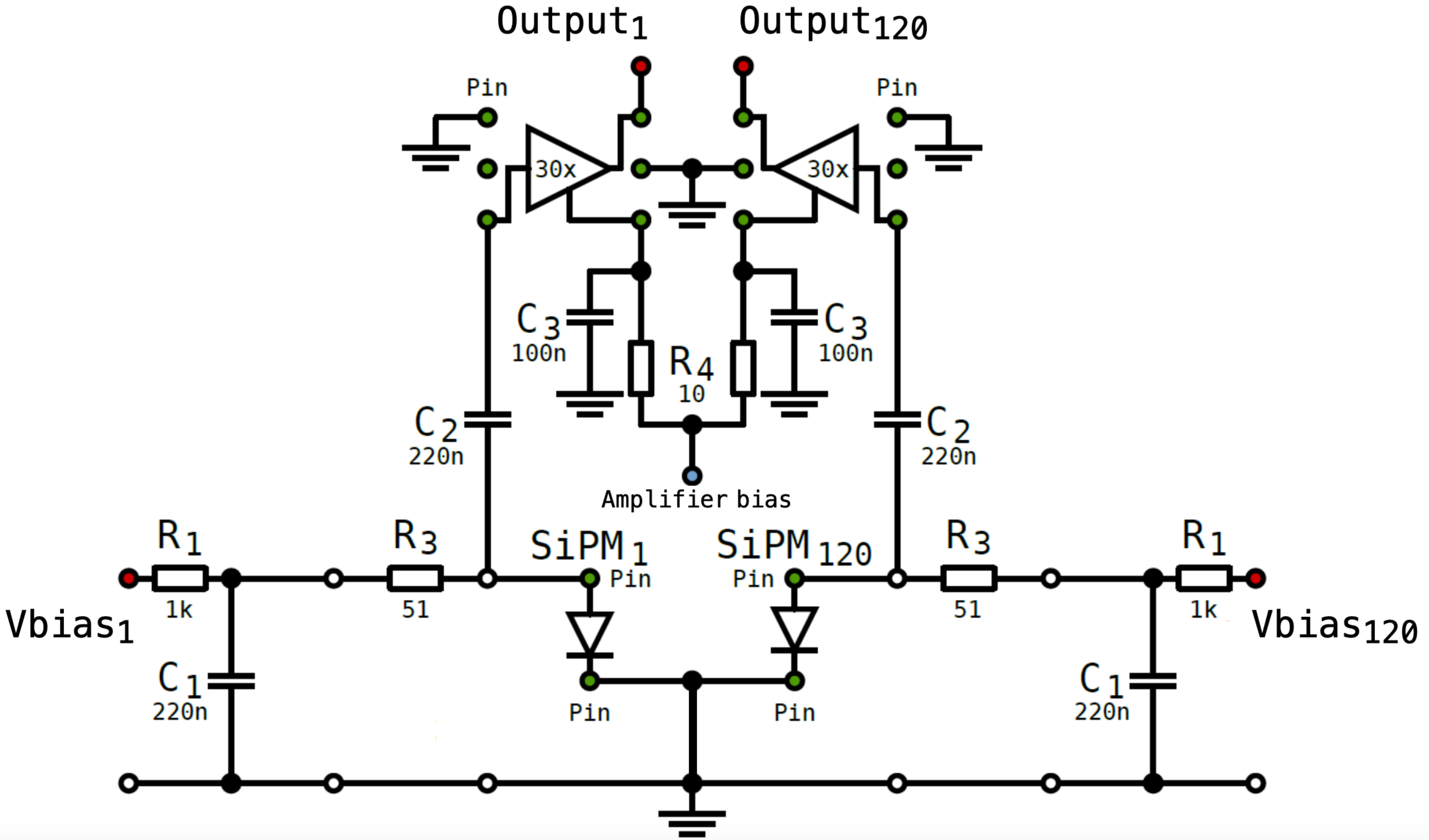}
    \caption{Circuit schematic of the bias and readout board. Output of 1 cell, marked as Output$_1$, is connected to the oscilloscope.}
    \label{fig:2ch-board}
\end{figure}

The readout board consists of two identical channels to bias and read out 1 cell and 120 cells separately. The circuit schematic of the board is shown in Figure \ref{fig:2ch-board}. The common cathode is connected to the ground. The anodes are AC-coupled to the separate amplifier circuits. In this study, the output signal of 1 cell was amplified by a Photonique SiPM Preamplifier AMP-0611 \cite{preamp}. The second channel was always operated at a 35~V bias voltage to keep 120 cells below the breakdown and avoid their impact on 1 cell via the common cathode or capacitive coupling from neighbour cells.

Measurements were carried out for one non-irradiated device and three devices irradiated by neutrons to different fluences ($\Phi$=2e12, 1e13, 5e13 cm$^{-2}$). 

The neutron irradiations were performed at room temperature without applied bias at the TRIGA Research Reactor of the JSI, Ljubljana. The samples were transported cold to Hamburg after irradiation and stored in a refrigerator
at -30$^{\circ}$C. No annealing was applied to the samples before measurement.

The data acquisition was synchronized with the laser trigger, its repetition rate was 1 MHz.

Waveforms of 400~ns length were recorded for 8-12 voltage settings above the breakdown voltage for each device with and without illumination. A non-irradiated device was used as reference, and was measured before and after each measurement of the irradiated devices to monitor the stability of the setup.

\section{Data analysis method}
\label{sec:method}
The acquired waveforms were processed using programs written in C++ and Python. Additionally, ROOT framework \cite{root} and GSL library \cite{gsl} were integrated into the analysis, which includes processing the raw data from the oscilloscope, SiPM pulse finding and validation.

\begin{figure}[h]
  \includegraphics[width=0.9\linewidth]{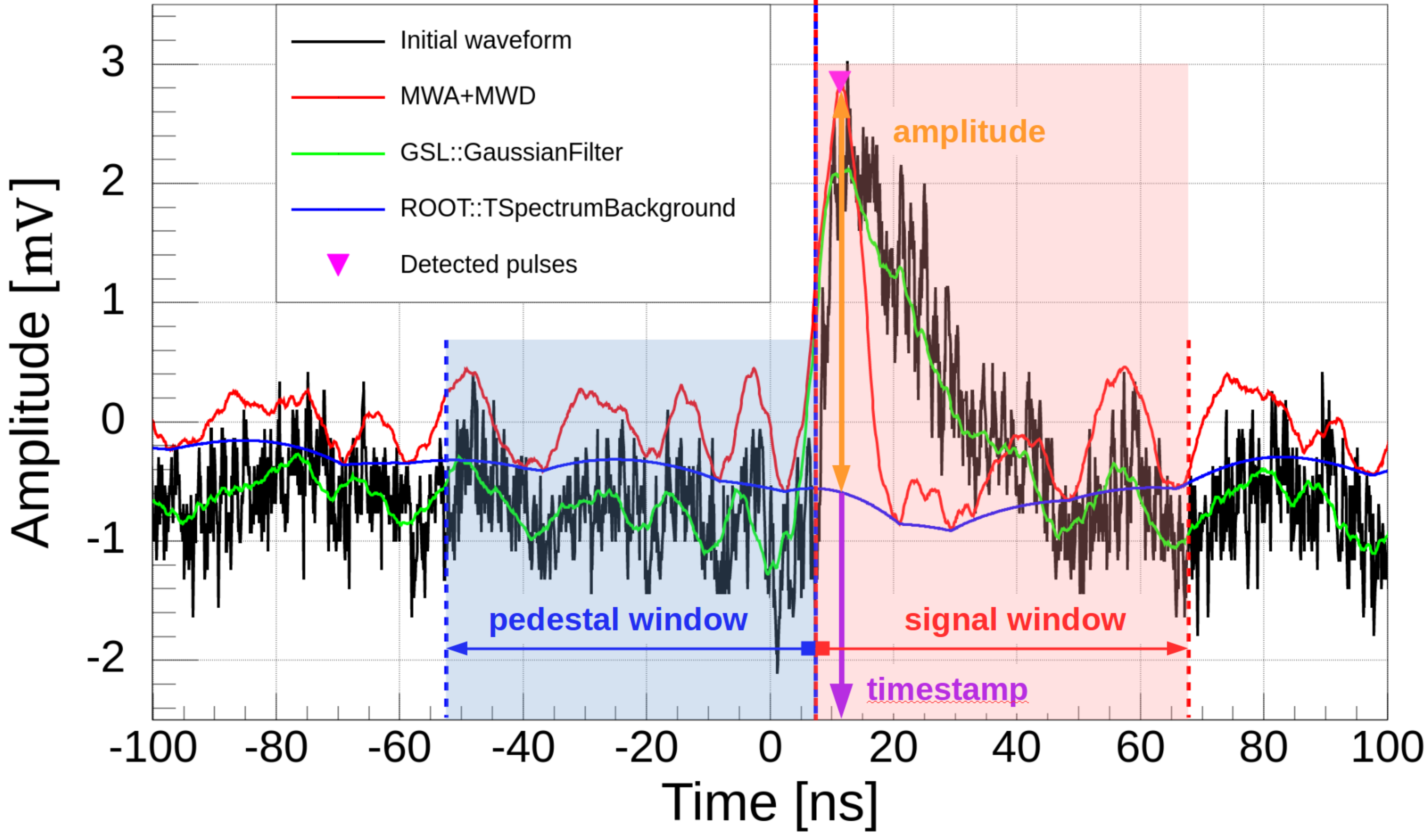} 
  \includegraphics[width=0.9\linewidth]{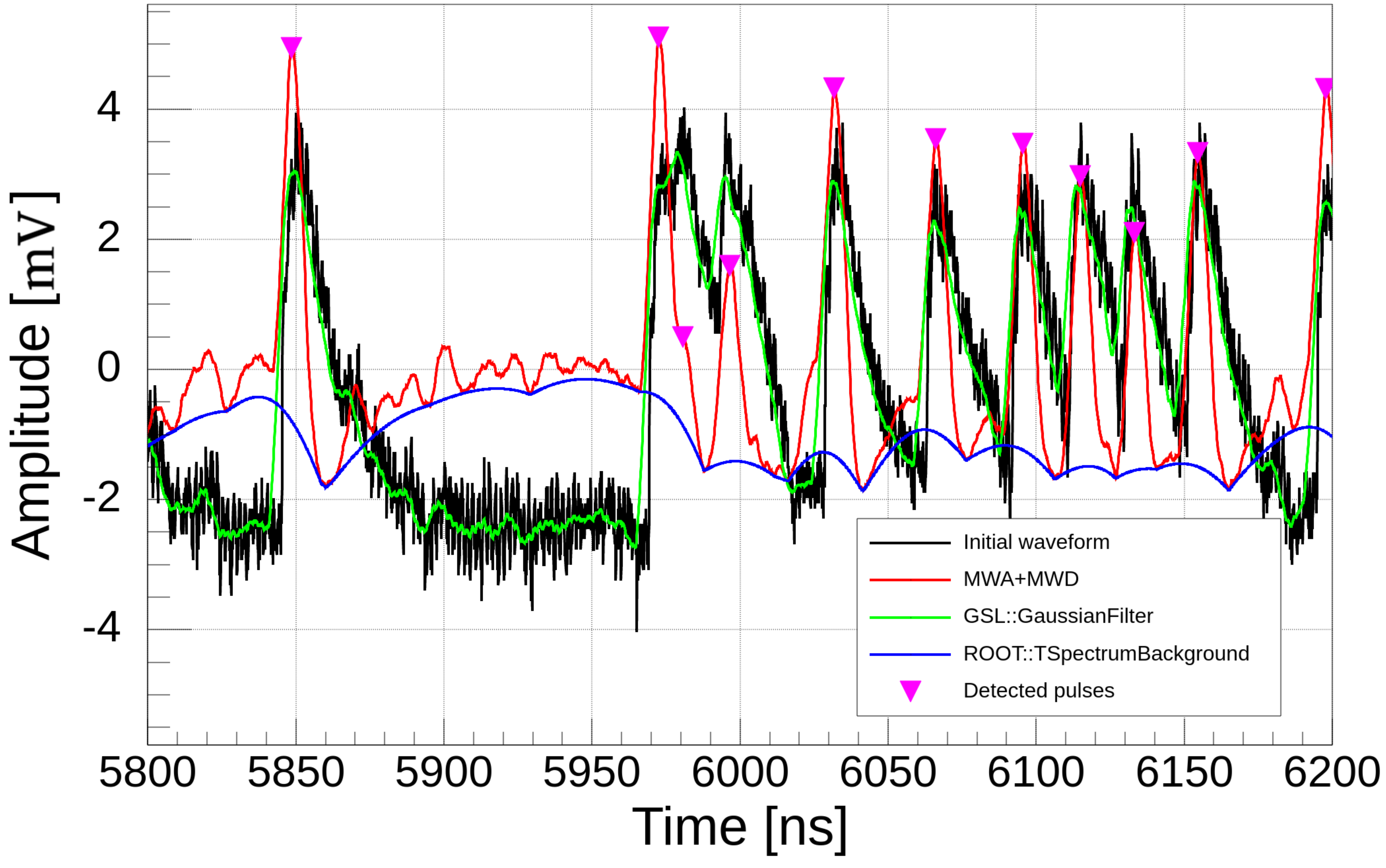}
  \centering
  \caption{Example of the pulse recognition algorithm (top) and its use on the sample irradiated to $\Phi$ = 5e13~cm$^{-2}$ fluence, measured at  $V_{ov}$ = 5.8~V, at -30$^{\circ}$C (bottom).}
  \label{fig:5e13_pulse_recog}
\end{figure}

The pulse finding algorithm combines
a moving window average (MWA) and moving window differential (MWD) method \cite{STEIN1996141}, 
with a moving window searching (MWS) algorithm. 
An example of the initial waveform and its transformations using the pulse finding algorithm are shown in Figure~\ref{fig:5e13_pulse_recog}.

The goals for the MWA+MWD method (red curve) are to reduce the electronics noise, and simultaneously to make SiPM pulses shorter to minimize their overlap in time. 
The MWS algorithm  is used to find the position in time of the amplitude maxima in the waveform. 
A search window of fixed width runs over each waveform measurement (amplitude vs time), compares the amplitude at each measured point in time with the amplitudes of N previous and N later measurements to find a maximum. The definitions of the time windows are reported in Table~\ref{table:algo_param}. The maximum amplitude within this window is tagged as probable SiPM pulse for the further analysis, and stored together with its corresponding time stamp.

The final step is to validate tagged pulses. This is required due to the presence of high frequency pick-up noise even on the filtered waveforms. 
For this the initial waveform is processed in parallel with two algorithms: Once with a Gaussian filter of 5.0~ns width to smooth the amplitude and remove the effect of high frequency pick-up noise (green curve), and once using the Sensitive Non-linear Iterative Peak Clipping algorithm (blue curve) to estimate the background.
The difference between the two transformed waveforms for the timestamp of the tagged pulses is compared to a threshold. If it exceeds the threshold, the tagged pulse is retained as SiPM signal. 
For each data set the threshold value that minimizes the number of fake pulses was iteratively found.

\begin{figure}[h]
  \includegraphics[width=0.9\linewidth]{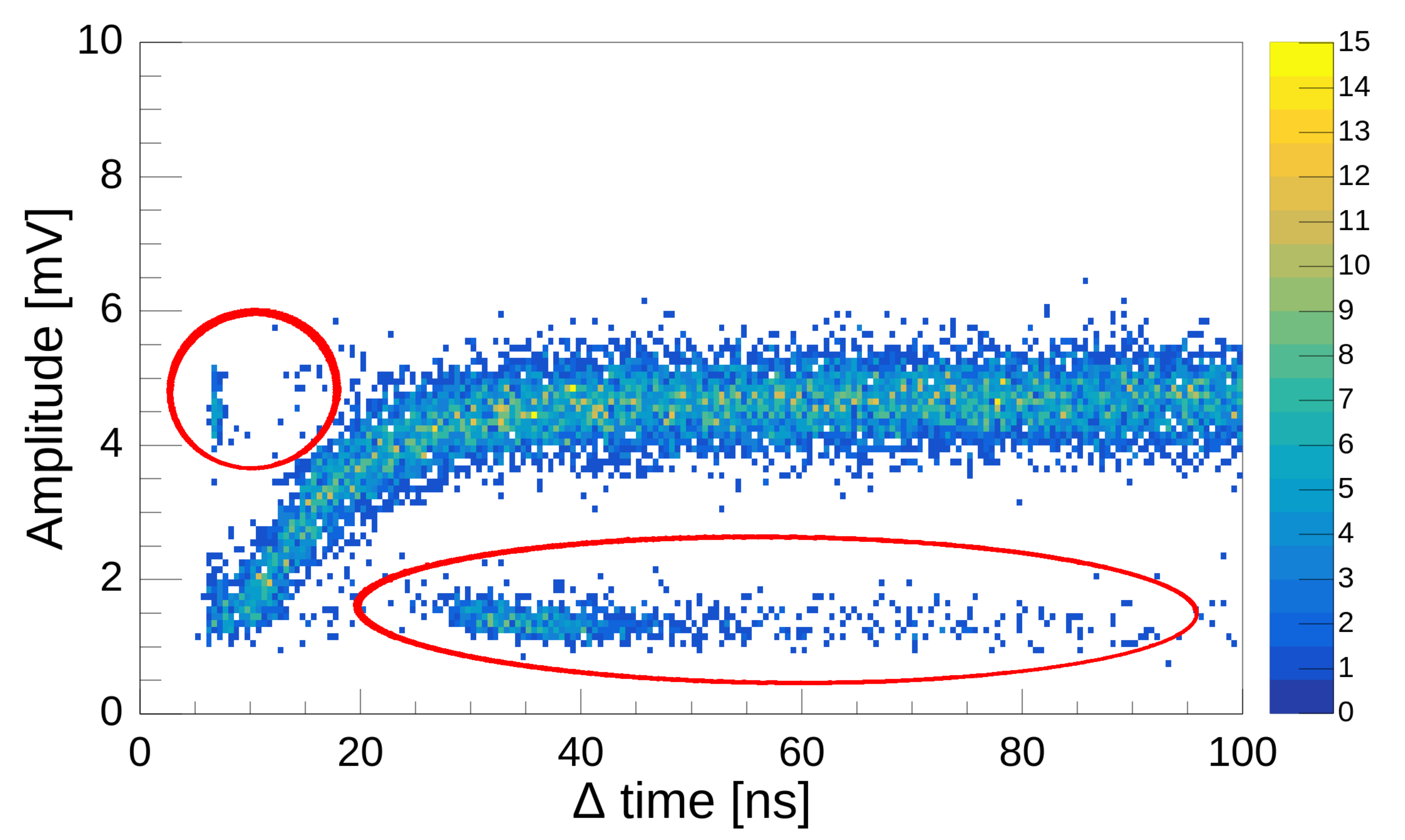}
  \includegraphics[width=0.9\linewidth]{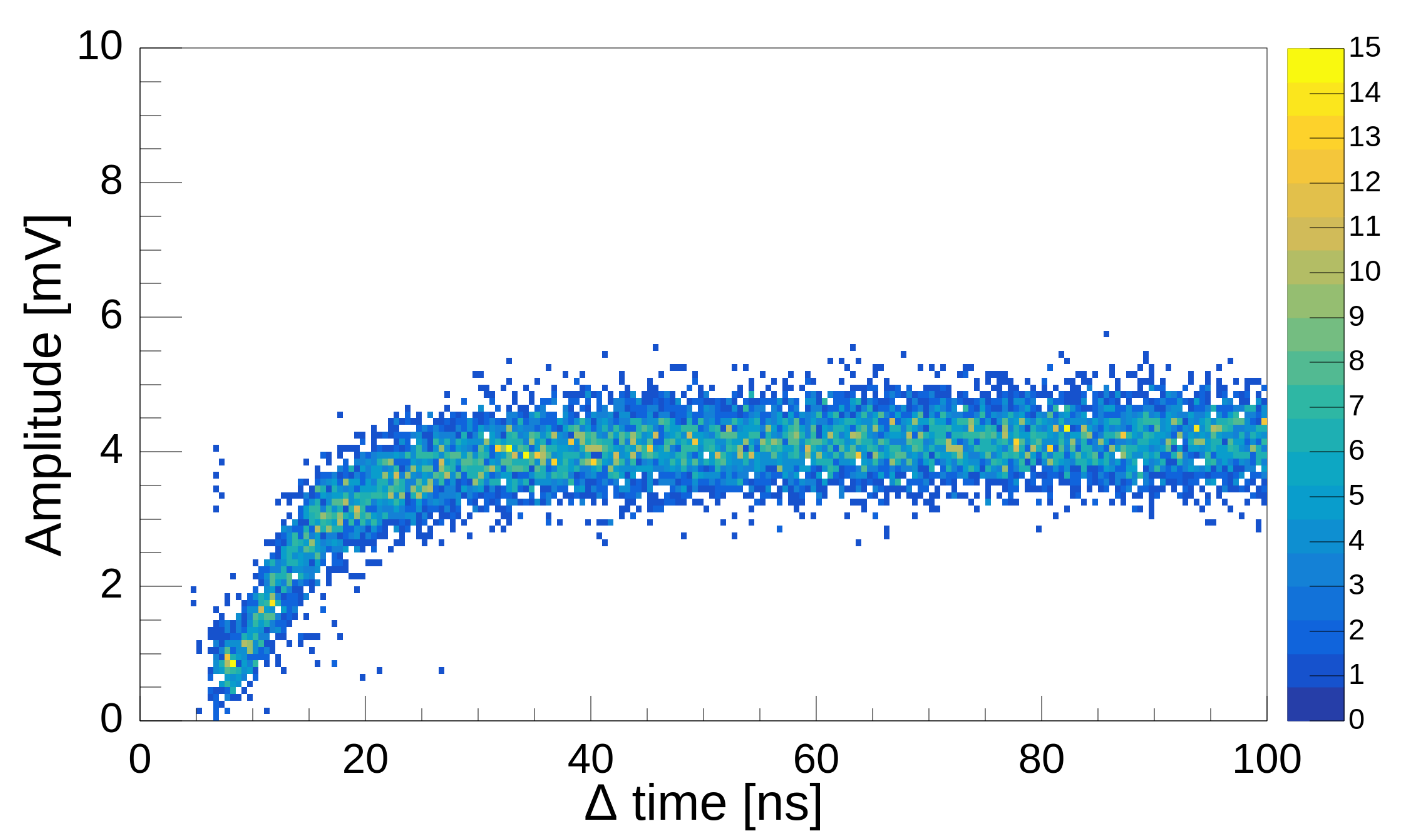}
  \centering
  \caption{Example of recovery time plots for the single cell sample irradiated to $\Phi$ = 2e12~cm$^{-2}$ fluence, measured at  $V_{ov}$ = 6.1~V, at 25$^{\circ}$C in dark conditions with sub-optimal algorithm parameters and fake pulses clusters marked in red (top) and with optimized parameters (bottom).}
  \label{fig:recoverytime}
\end{figure}

The qualitative and quantitative analysis of the pulse finding and validation algorithm was based on the SiPM recovery plot of which an example is shown in Figure~\ref{fig:recoverytime}. Here the amplitude of each pulse is plotted versus the time difference of the current to the previous pulse. 

\begin{table}[h!]
  \begin{center}
    \begin{tabular}{c|c|c|c} 
      \hline
      \textbf{Window} & \textbf{Length, ns} & \textbf{$t_{start}$} & \textbf{$t_{stop}$} \\
      \hline
      $\Delta t_{MWA}$ &  5.5 & $t_i$      & $t_i + \Delta t_{MWA}$   \\
      \hline
      $\Delta t_{MWD}$ &  6.2 & $t_i$      & $t_i + \Delta t_{MWD}$   \\ 
      \hline
      $\Delta t_{MWS}$ & 3.1  & $t_i - N $ & $t_i + N $              \\ 
      \hline
    \end{tabular}
    \end{center}
  \caption{Definition of the time windows for the methods applied in the pulse finding algorithm. The time length of the window is given by the number of 0.1 ns spaced measurements points considered.}
  \label{table:algo_param}
\end{table}

\subsection{Pulse counting}
\label{sec:pulse}
After applying the pulse finding algorithm the number of SiPM pulses in each waveform is counted. In the absence of external illumination this number is proportional to the dark count rate (DCR) of the SiPM, or of the single cell.   
For example, it was measured that a non-irradiated single cell has a dark count rate  of the order of 10~Hz at the temperature 25~$^{\circ}$C for $V_{ov} =$~4~V overvoltage. In case of multi-cell SiPM one measures the sum of from all cells DCRs.
Hamamatsu provides a typical DCR value of 120~kHz measured at 25~$^{\circ}$C at $V_{ov} =$~4~V for the 1.3x1.3~mm$^{2}$ size non-irradiated device having 7284 cells \cite{HPKS14160} which gives an initial $DCR_0$ of about 16~Hz for a single cell and is compatible with our data. 

After irradiation the single cell DCR is increasing above the initial value as a linear sum of two components $DCR_{\Phi} = DCR_0 + DCR_{irr}$ and for fluence $\Phi$ = 2e12~cm$^{-2}$ at the temperature 25~$^{\circ}$C and 4~V overvoltage the DCR reaches 2~MHz. At this fluence the term $DCR_{irr}$ originated from the radiation damage dominates. 
Under the assumption that all cells are equally damaged, one can use measurements on the single cell to infer $DCR_{\Phi}$ of an SiPM with the same cell design by linearly scaling with the number of cells. 

\begin{figure}[h]
  \includegraphics[width=0.95\linewidth]{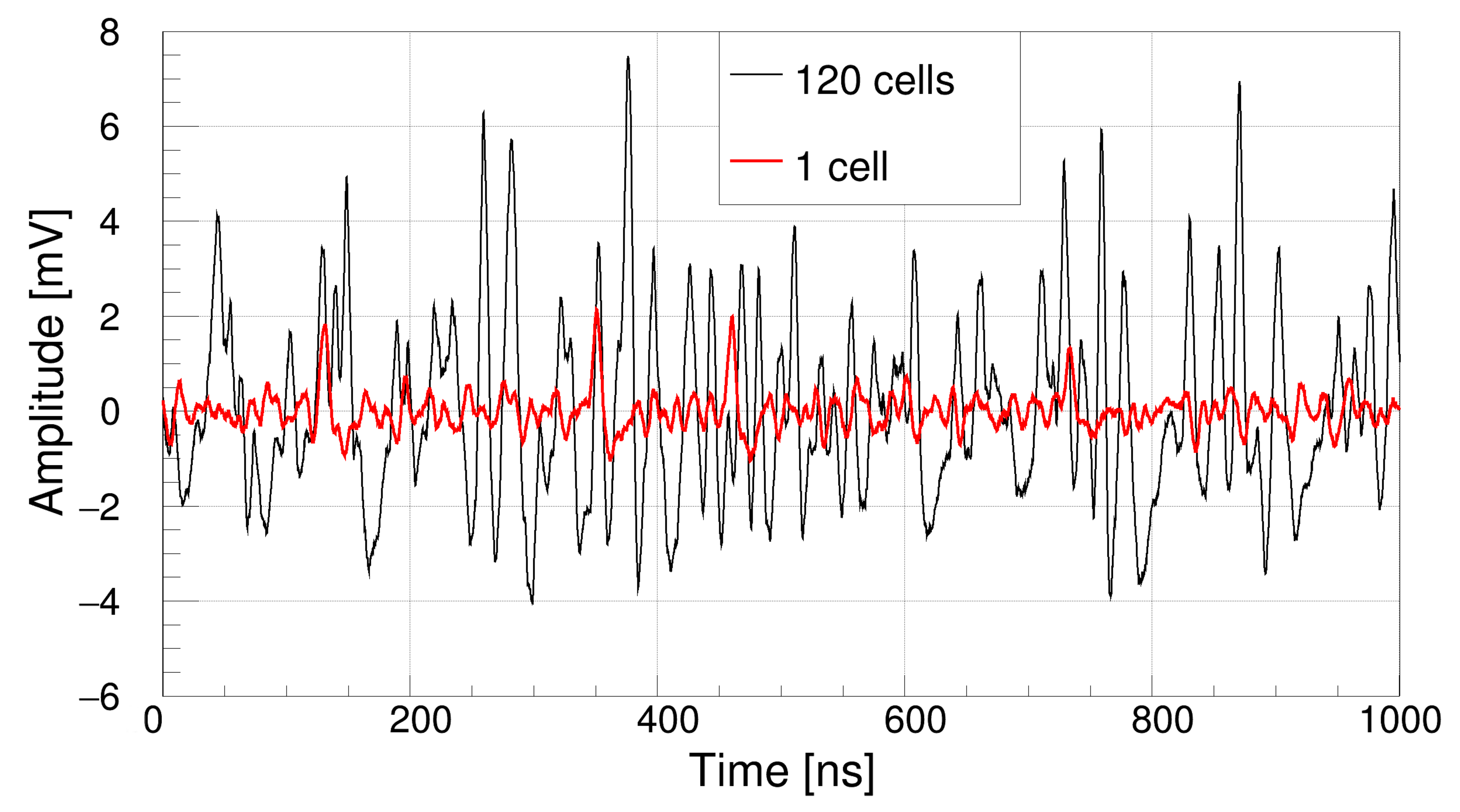} 
  \includegraphics[width=0.95\linewidth]{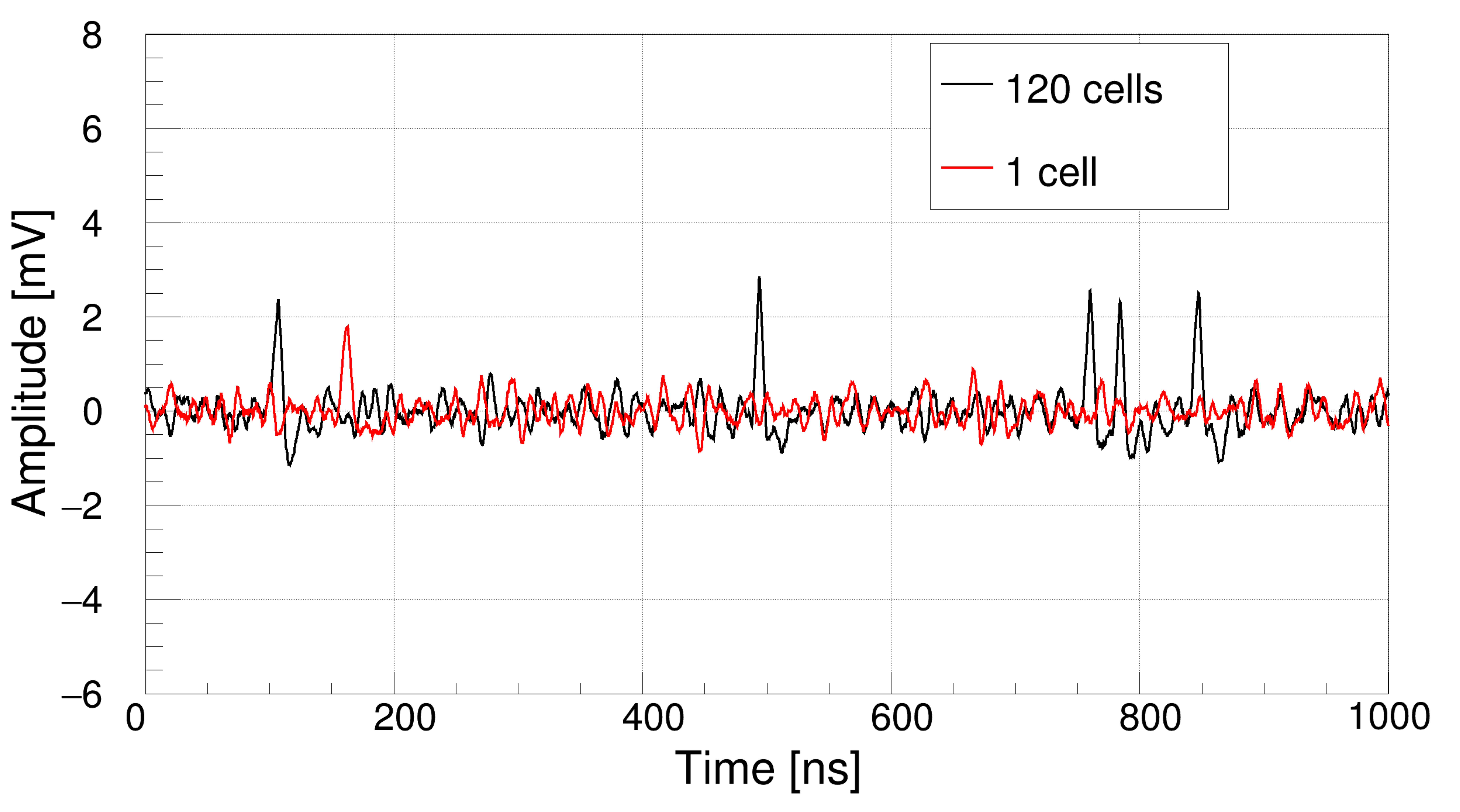} 
  \centering
  \caption{Example of filtered waveforms acquired for the sample irradiated to $\Phi$ = 2e12~cm$^{-2}$ fluence, under  $V_{ov}$ = 4.1~V, at 25~$^{\circ}$C in dark conditions (top) and for the non-irradiated sample using the same overvoltage and temperature with a continuous LED light illumination (bottom)}
  \label{fig:1-vs-120_wfs}
\end{figure}

\begin{figure}[h]
  \includegraphics[width=0.95\linewidth]{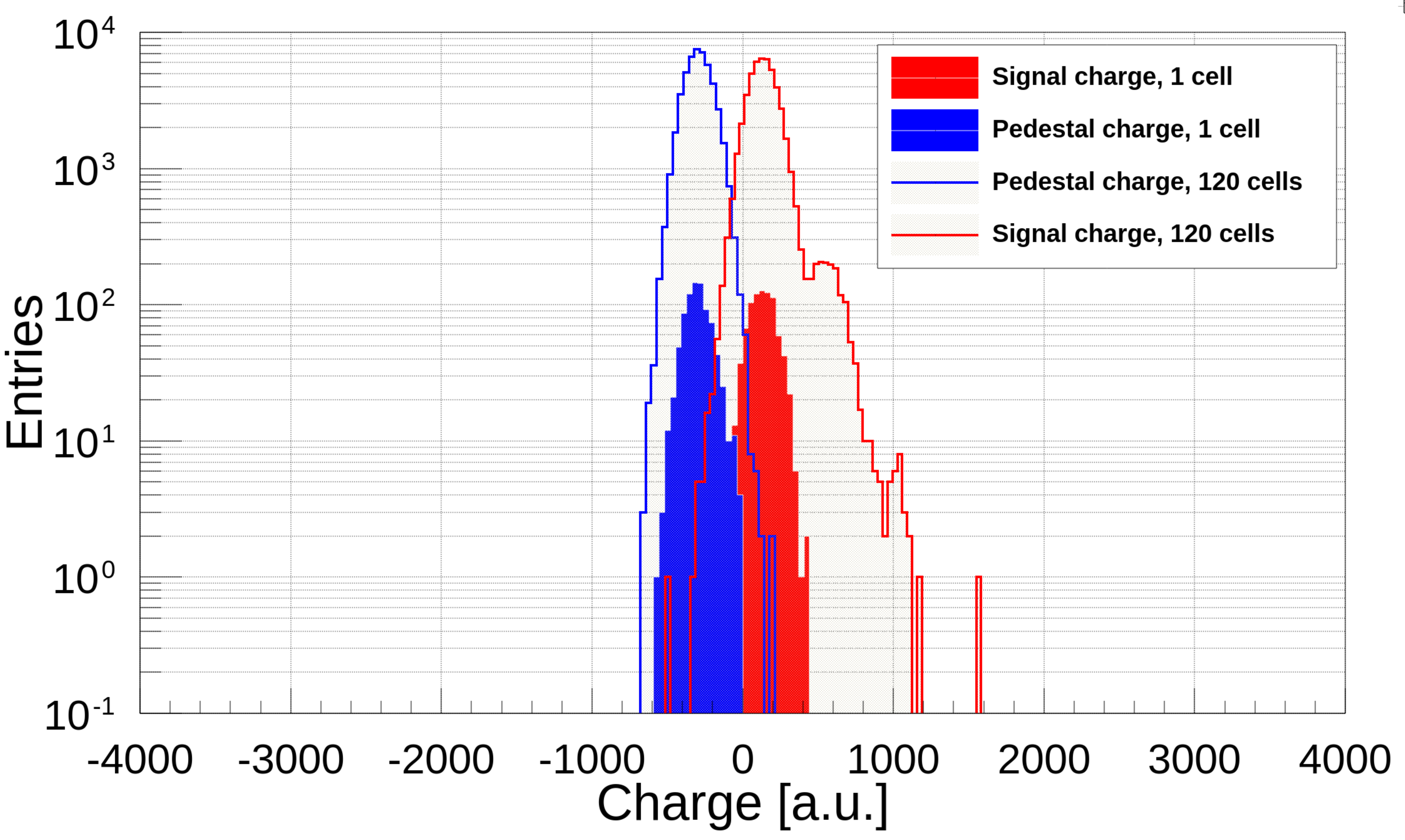} 
  \includegraphics[width=0.95\linewidth]{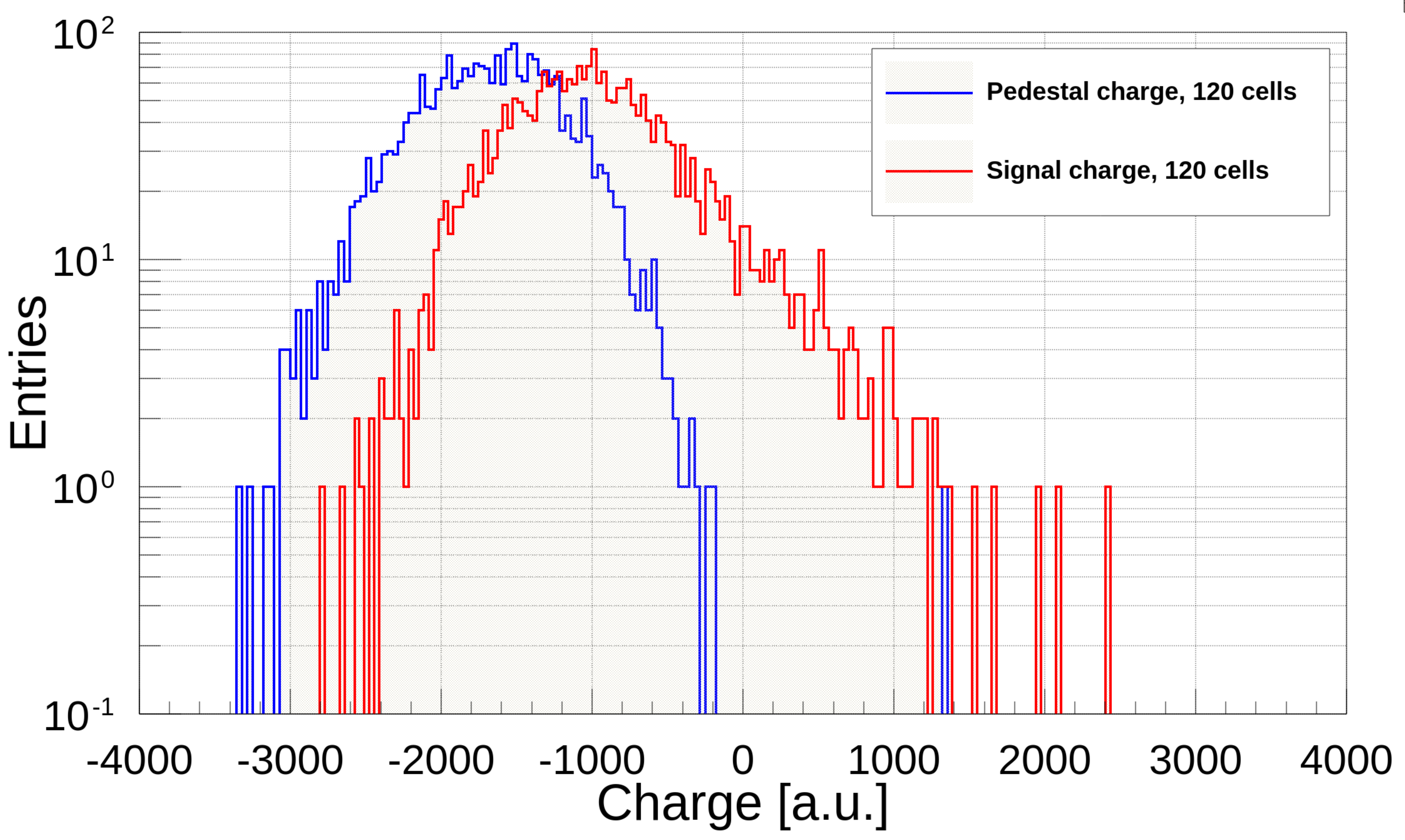}
  \centering
  \caption{Charge spectra for the data from 120 cells readout measured at $V_{ov}$ = 4.1~V, at 25~$^{\circ}$C for non-irradiated (top) and irradiated to $\Phi$ = 2e12~cm$^{-2}$ fluence (bottom) samples. On the top histogram a charge spectrum for a single cell readout for the same conditions is provided.}
  \label{fig:1-vs-120_spectra}
\end{figure}

For the 120-cells SiPM irradiated with fluence $\Phi$ = 2e12~cm$^{-2}$ for 4~V overvoltage and T=25~$^{\circ}$C a $DCR_{\Phi}$ around 240~MHz is expected. The filtered waveforms of a single cell and a 120-cells SiPM for these conditions are displayed in the top plot of Figure~\ref{fig:1-vs-120_wfs}. The 120-cells SiPM does not show single cell response. For comparison the filtered waveforms from a non-irradiated single cell and 120-cells SiPM are presented on the bottom plot of  Figure~\ref{fig:1-vs-120_wfs}. Here, pulses induced by low intensity continuous LED illumination are clearly visible in both waveforms. This is also confirmed using charge spectra histograms for 60 ns integration time - the ones obtained from the 120-cells SiPM do not provide SPE resolution already at the fluence of $\Phi$ = 2e12~cm$^{-2}$ while the non-irradiated sample has clear SPE peaks as shown on Figure~\ref{fig:1-vs-120_spectra}. On the top part of the figure a charge spectrum from 1 cell is also provided. 

In order to minimize DCR, SiPM are actively cooled. At the T=-30~$^{\circ}$C the DCR of a single cell with 4~V overvoltage for the fluence $\Phi$ = 2e12~cm$^{-2}$ is about 0.1~MHz, while for the fluence $\Phi$ = 5e13~cm$^{-2}$ the DCR is about 10~MHz.

\subsection{Charge integration}
\label{sec:charge}

\begin{figure}[h]
  \includegraphics[width=0.9\linewidth]{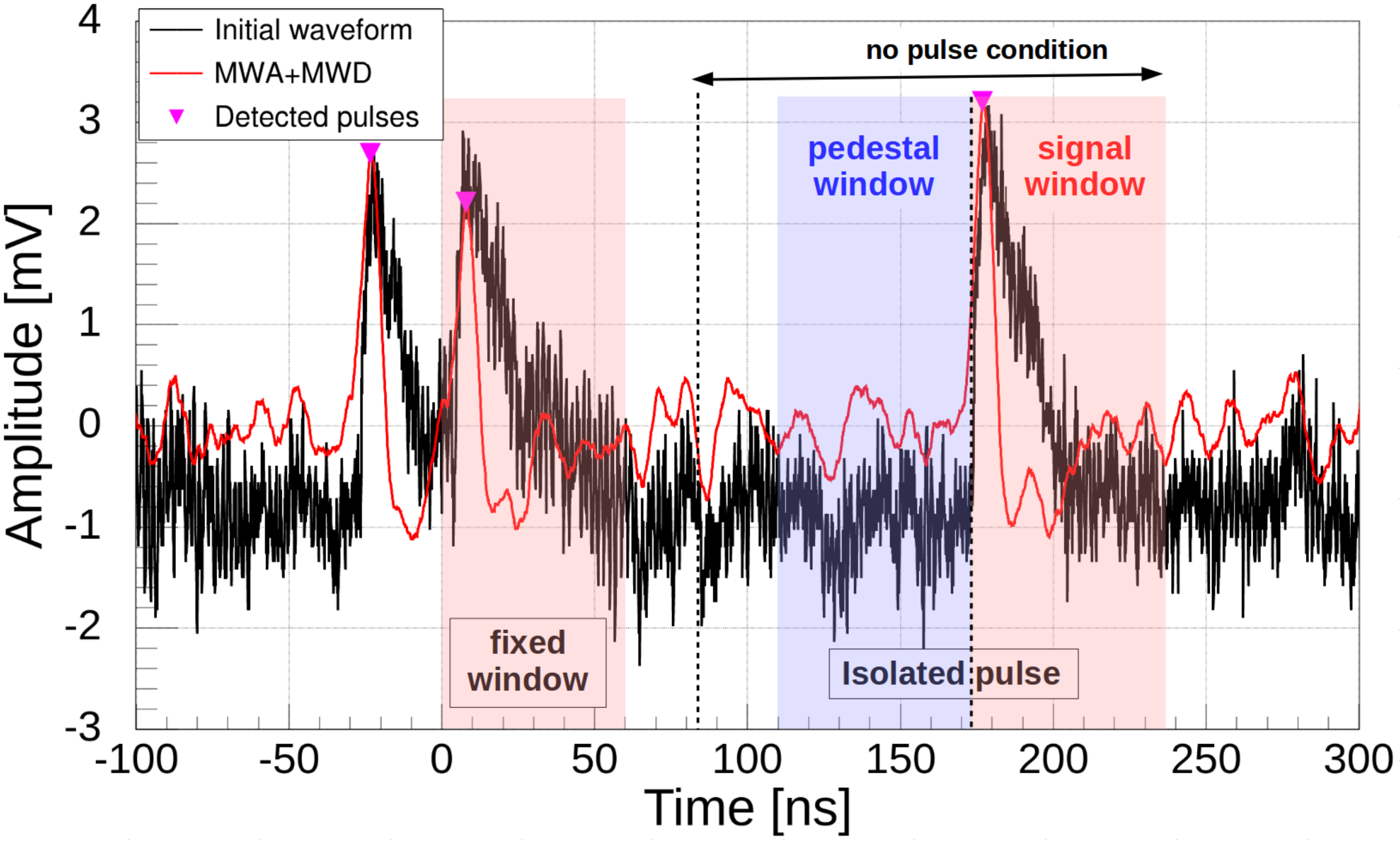}
  \centering
  \caption{Schematic representation of the charge integration using a fixed window and isolated pulses for the sample irradiated to $\Phi$ = 5e13~cm$^{-2}$, measured at  $V_{ov}$ = 4.0~V, at -30$^{\circ}$C.}
  \label{fig:isol_fixed}
\end{figure}

In this work we compared two approaches for the pulse charge integration. 
Both methods are schematically presented in Figure~\ref{fig:isol_fixed}.

The first one uses a gate of given width synchronised with the trigger. 
In this case no pulse recognition is required. This fixed window (FW) technique is useful for the data acquired with the laser, where the time of the expected pulse is known.

In the second method, we detect all the pulses on every waveform and select only isolated pulses (IP). The time isolation criteria requires no additional pulses in the 90~ns window before the pulse time stamp, and only one pulse in 60~ns time window after the time stamp. The former is sufficient for the cell to fully recover after a discharge, and is chosen based on the recovery time plot in Figure~\ref{fig:recoverytime}. After event selection, pedestal and signal charges are obtained integrating the non-filtered waveforms separately in gates of equal length before and after the pulse starting point. This is defined as the time stamp of the tagged pulse minus 30 points, or 3.0~ns. 

\section{Results}
\label{sec:results}
\label{sec:gain}
\begin{figure}[h]
  \includegraphics[scale=0.1]{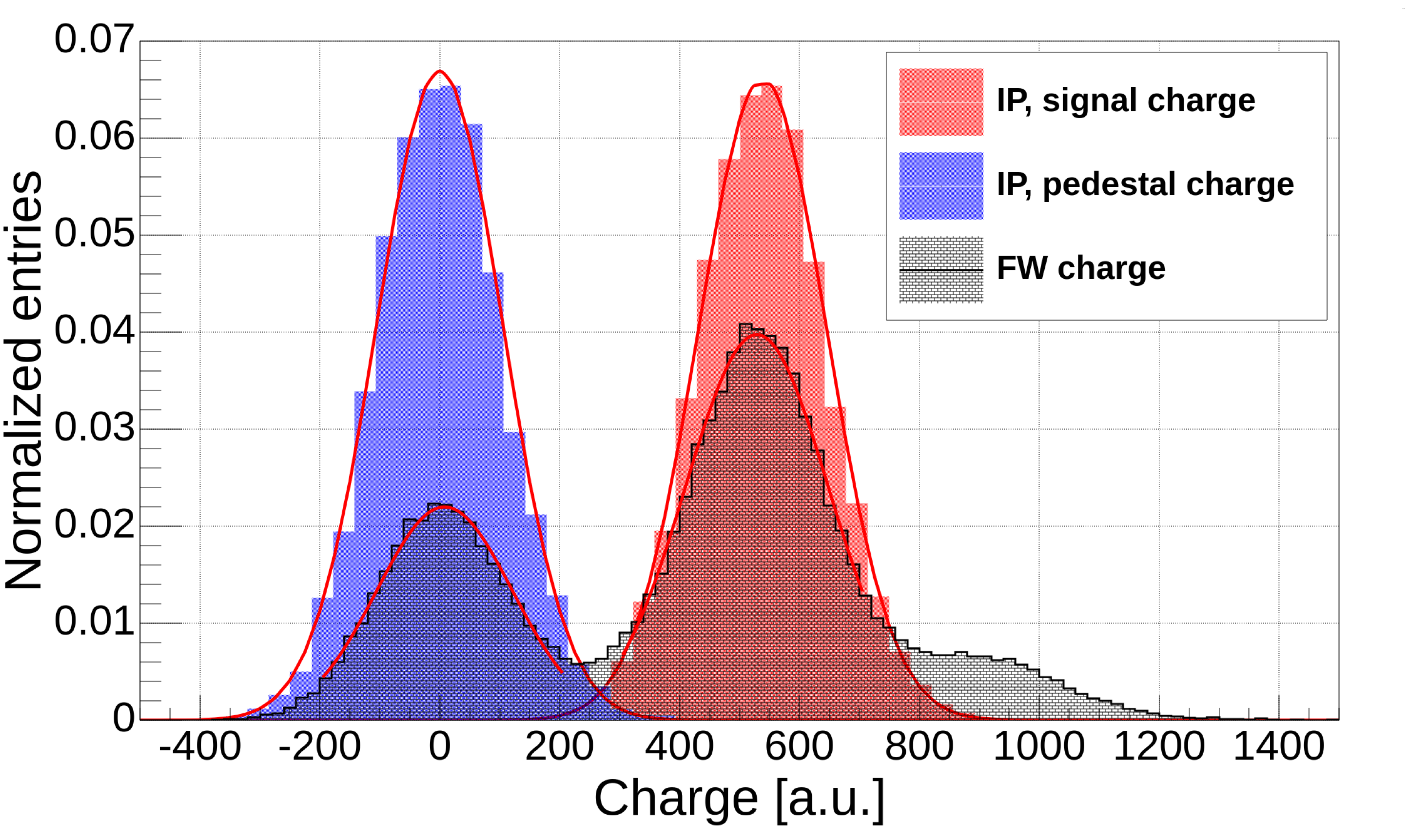}
  \centering
  \caption{Histograms for the charge integrated in 60~ns gate for the laser data with pedestals shifted to 0 by 464.3 and 454.9 for IP and FW methods, respectively, for the sample irradiated to $\Phi$ = 5e13~cm$^{-2}$, measured at  $V_{ov}$ = 3.2 V, at -30 $^{\circ}$C.}
  \label{fig:charge_histograms}
\end{figure}

The gain of the DUT is defined as the distance between the most probable values (MPV) of the pedestal and of the first single photo-electron peak. For a single-cell device, no higher order peak is observed as expected.
An example of such histograms normalized to the total number of entries is shown in Figure~\ref{fig:charge_histograms} for the two charge integration methods. The pedestal peaks are  shifted to zero by subtracting the respective offsets. 
Using the FW method a shoulder is visible on the right of the first single photo-electron peak.
This is the case only for highly irradiated samples. It is caused by partial integrating a second pulse within the gate, because of a very high $DCR_{\Phi}$. A charge larger than that of a single pulse can be recorded in this case. This effect is reduced if the integration gate is shortened.\\

\begin{table}[h!]
  \begin{center}
    \begin{tabular}{c|c|c|c} 
      \hline
      \textbf{Data} & \textbf{Entries, N} & \textbf{Gaussian $\mu$} & \textbf{Gaussian $\sigma$}\\
      \hline
      FW, pedestal & 90000 & 0.0$\pm$0.9 & 108.9$\pm$0.9 \\
      \hline
      FW, signal & 90000 & 525.55$\pm$0.63 & 121.4$\pm$0.7 \\ 
      \hline
      IP, pedestal & 10375 & 0.0$\pm$1.0 & 106.0$\pm$0.8 \\ 
      \hline
      IP, signal & 10375 & 538.51$\pm$1.06 & 107.9$\pm$0.8 \\ 
      \hline
    \end{tabular}
  \end{center}
  \caption{Summary of histograms and Gauss fit statistics presented in Fig.~\ref{fig:charge_histograms}.}
  \label{table:1}  
\end{table}

\begin{figure}[h]
  \includegraphics[width=0.95\linewidth]{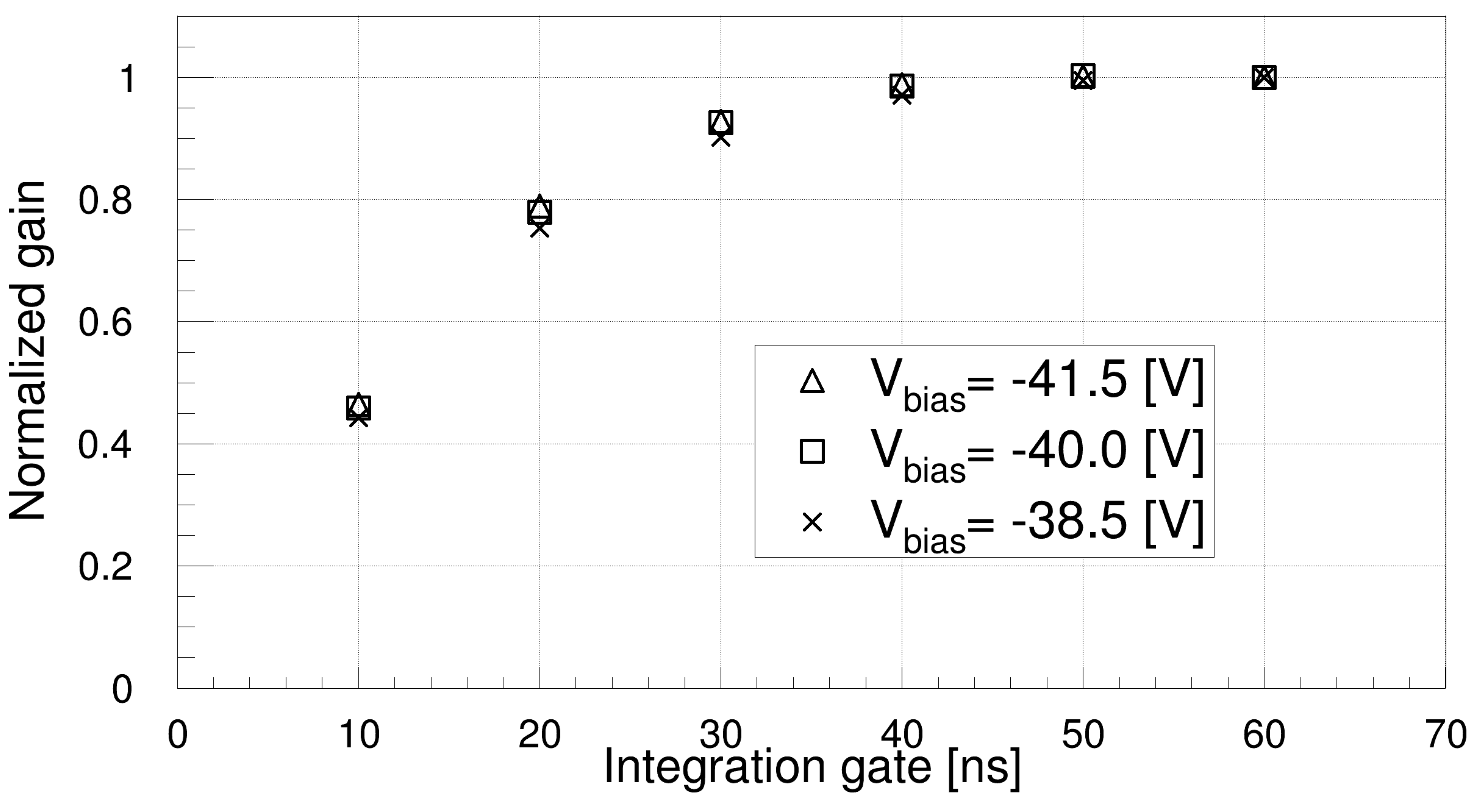}
  \centering
  \caption{Effective gain $G^*$ as a function of the integration gate for the sample irradiated to $\Phi$ = 5e13~cm$^{-2}$, at -30 $^{\circ}$C. Only for a gate length larger than 50~ns $G^* = G$.}
  \label{fig:normalized_gain}
\end{figure}

The gate length is selected such that the full pulse charge is integrated. For a shorter gate length only an effective gain, $G^*$ is obtained which is smaller than the true gain of the device, as demonstrated by the integration gate length scan in Figure~\ref{fig:normalized_gain}. For a gate length of 20~ns the effective gain $G^*$ is 20\% smaller than the gain. 

\begin{figure}[h]
  \includegraphics[width=0.9\linewidth]{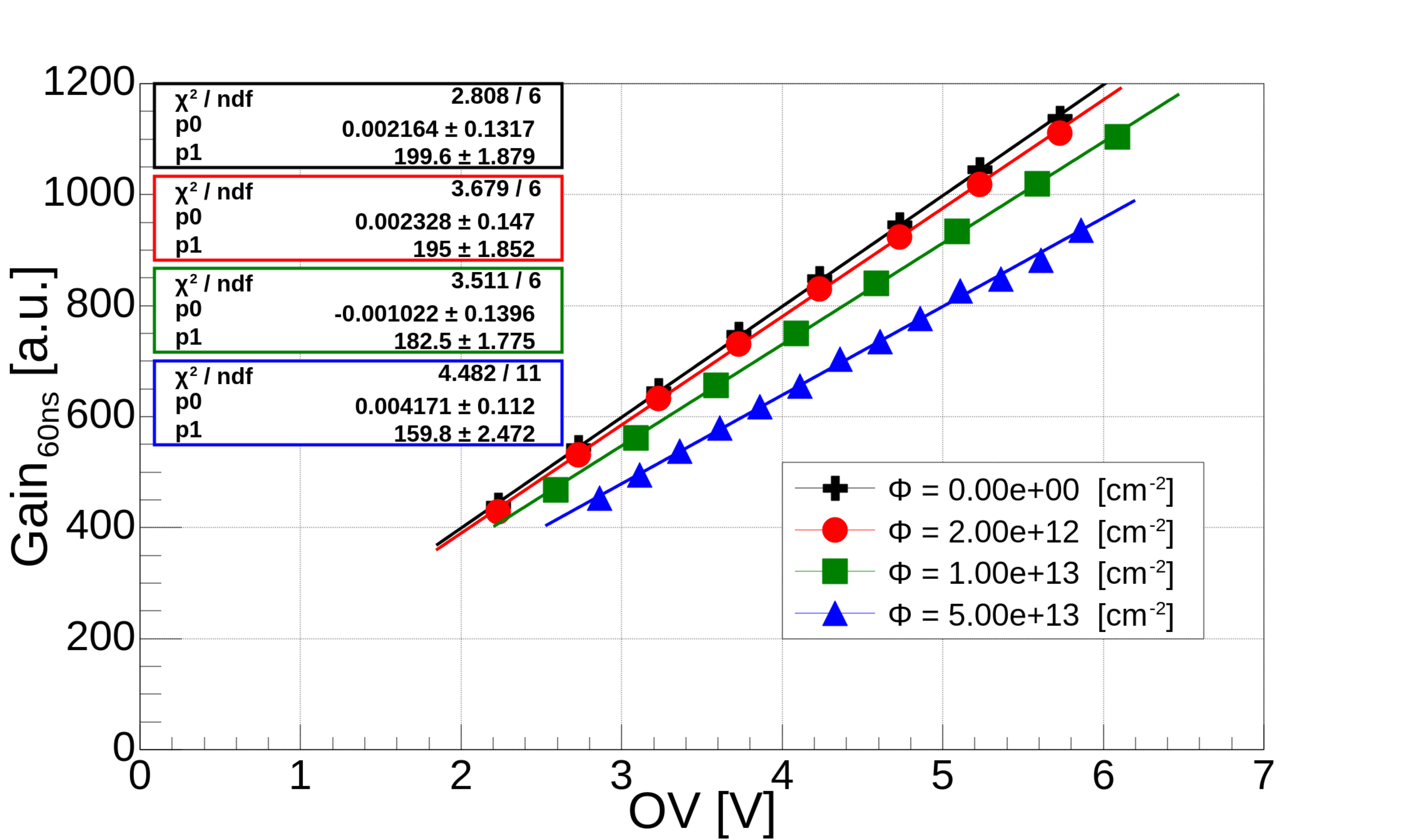}
  \includegraphics[width=0.9\linewidth]{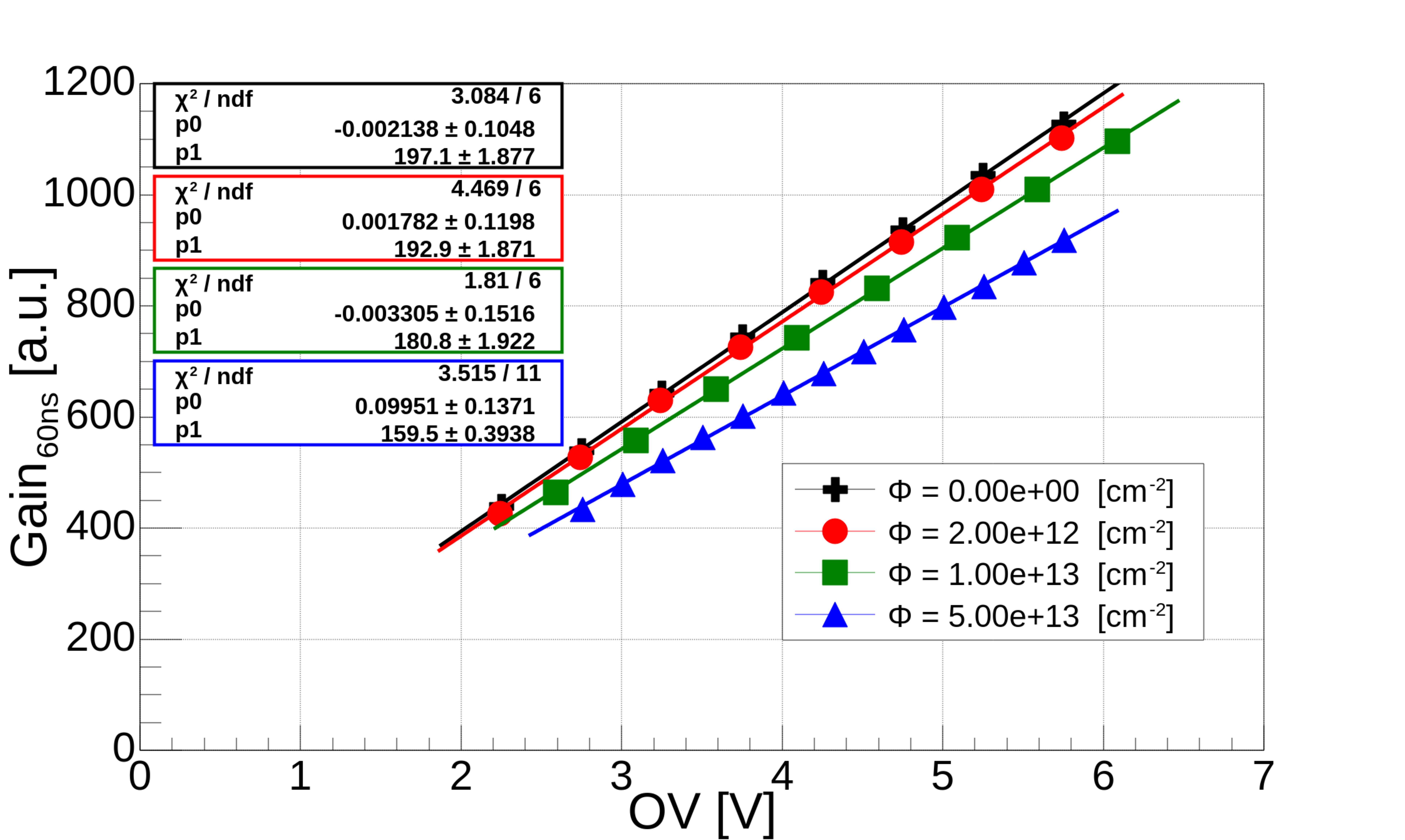}
  \centering
  \caption{Gain versus overvoltage for the samples measured at -30 $^{\circ}$C obtained integrating laser induced pulses in a 60~ns gate for IP (top) and FW (bottom). Data for non-irradiated (black crosses), $\Phi$ = 2e12~cm$^{-2}$ (red circles), $\Phi$ = 1e13~cm$^{-2}$ (green squares) and $\Phi$ = 5e13~cm$^{-2}$ (blue triangles) samples are shown. For each data set a line of the same color representing linear fit using $y=p0+p1*x$ is given, where $p0=-V_{off}*dGdV$ and $p1=dGdV$. Turn-off voltages for each sample are given in Table.\ref{table:2}.}
  \label{fig:turnoff_voltages}
\end{figure}

The turn-off voltage $V_{off}$ is extracted as a fit parameter of the function $Gain(V) = dGdV * (V – V_{off})$ used for the data approximation of the gain versus voltage scan measurements.
Figure~\ref{fig:turnoff_voltages} presents the linear regressions for the non-irradiated and irradiated samples at -30 $^{\circ}$C . It should be noted that linear dependence of gain on overvoltage shows that there is no effect of sensor self-heating due to Geiger discharges even for heavily irradiated samples.
The value of $V_{OV}$ on the x-axis is calculated using $V_{off}$ from the fit. 

The extracted parameters are summarized in Table~\ref{table:2}. 
For the same overvoltage, a few percent reduction of gain is visible at $\Phi$ = 1e13 cm$^{-2}$, which increases to 19\% at $\Phi$ = 5e13 cm$^{-2}$. At this fluence the turn-off voltage increases by about half of a Volt.
The results obtained using the two selection methods are compatible within errors. 
The one obtained using isolated pulses charge integration are used in Figure~\ref{fig:normalized_gain}.
If the integration gate is reduced to 20~ns, the effective gain before irradiation is 20\% lower than $G$ and reduces by 7\% at $\Phi$ = 5e13 cm$^{-2}$.
 
\begin{table*}[t]
  \begin{center}
    \begin{tabular}{c|c|c|c} 
      \hline
      \textbf{$\Phi$, cm$^{-2}$} & \textbf{Slope} & \textbf{V$_{off}$, V} & 
      \textbf{$G/G_0$} \\
      \hline
      \multirow{2}{*}{0e00} & 199.6$\pm$1.88  & 35.27$\pm$0.04 & 1.00 \\
      & 197.1$\pm$1.88 & 35.25$\pm$0.04 & 1.00 \\
      \hline
      \multirow{2}{*}{2e12} & 195.0$\pm$1.85 & 35.27$\pm$0.04 & 0.98 \\
      & 192.9$\pm$1.87 & 35.26$\pm$0.04 & 0.98 \\
      \hline
      \multirow{2}{*}{1e13} & 182.5$\pm$1.78 & 35.41$\pm$0.04 & 0.91  \\
      & 180.8$\pm$1.92 & 35.41$\pm$0.04 & 0.92 \\
      \hline
      \multirow{2}{*}{5e13} & 159.8$\pm$2.47 & 35.64$\pm$0.06 & 0.80 \\
      & 159.5$\pm$0.39 & 35.74$\pm$0.05 & 0.81 \\ 
      \hline
    \end{tabular}
  \end{center}
  \caption{Summary of the fit values extracted from Fig.~\ref{fig:turnoff_voltages}. 
  Slopes, V$_{off}$ and the gain ratio between the given sample $G$ and non-irradiated one $G_0$. For each fluence, the results for isolated pulses (top rows) and fixed window (bottom rows) techniques are given. All parameters are quoted for the temperature of -30~$^{\circ}$C.}
  \label{table:2}  
\end{table*}

\section{Conclusions and outlook}
A single-cell SiPM was designed and produced by Hamamatsu for dedicated radiation damage studies. Measurements of such structure were not presented so far, and make it possible to investigate key properties of silicon photomultipliers and their dependence on radiation damages at the level of a single cell.
The structure has a 15~$\mu$m pitch and was irradiated with neutrons to $\Phi$ = [2e12, 1e13, 5e13] cm$^{-2}$. It has been measured for voltages between 2 and 6 V above the turn-off voltage, at -30~$^{\circ}$C, with and without laser light illumination. 
A method for the data analysis is developed, which includes initial waveform processing, SiPM pulse recognition and validation and pulse parameters calculation.
From the acquired data, the gain and turn-off voltage 
are extracted. 
A reduction of the gain by 19\% and an increase of $V_{off}$ by $\approx$0.5~V is observed after $\Phi$ = 5e13 cm$^{-2}$.   

The exact cause of the gain reduction after irradiation is not known and will require more in-depth investigations. Possible explanations for this effect could be: enhanced charge trapping or recombination in the active cell volume; 
reduction of effective cell volume due to formation of large size cluster defects by neutron irradiation; and
reduction of effective cell volume due to change of doping concentration. 
 
Studies of the introduction rate and macroscopic effects of CiOi and BiOi defects, as pointed out in \cite{moll2020}, may be relevant to understand this point further. Also TCAD-based simulations including a radiation damage model \cite{schwandt2018} may be helpful to understand which of the above mentioned effect dominates in the case of SiPMs.

The opportunity to study basic parameters of a SiPM single cell opens a new way to understand SiPM behaviour in various environments, especially after irradiation. Though, one critical aspect in using this method is the localisation of radiation-induced damage. It has been observed, see for instance \cite{Eugen2018:arxiv}, that radiation damage may produce local "hot spots", with size smaller than or similar to a cell. The probability that one of this hot spots is found in the single cell under study depends on the fluence and the cell size. The results presented in this paper could be subject to large fluctuations due to the presence of "hot spots". They should be compared to the average results of few cells in future studies.  
Next the fluence dependence of the difference between $V_{off}$ and $V_{bd}$ will be investigated, which requires dedicated low current measurements on the single cell.  

\section*{Acknowledgement}
The authors wish to thank HPK for providing the MPPC structure for these studies.
We thank Jens Schaarschmidt for the development of the readout and amplifier boards used in the setup. We also thank Arjan Heering and Yuri Musienko for organizing the irradiation of the devices. 
This work is supported by the Deutsche Forschungsgemeinschaft (DFG, German Research Foundation) under Germany's Excellence Strategy, EXC 2121, Quantum Universe (390833306).
The reported study was funded by RFBR and TUBITAK according to the research project 20-52-46005. 

\bibliographystyle{elsarticle-num}
\bibliography{article}

\end{document}